\documentclass[usenatbib]{mn2e}

\usepackage{epsfig}
\usepackage{amssymb}

\begin{document}

\title[SCUBA observations of HEROs in the SDF]{Submillimetre Constraints on Hyper-Extremely Red Objects in the Subaru Deep Field}
\author[Coppin et al.]{
\parbox[t]{\textwidth}{
\vspace{-1.0cm}
%Kristen Coppin$^{1}$, Mark Halpern$^{1}$, Douglas Scott$^{1}$, 
%Gaelen Marsden$^{1}$, Fumihide Iwamuro$^{2}$, Toshinori Maihara$^{3}$, Kentaro Motohara$^{4}$, Tomonori Totani$^{5}$}
Kristen Coppin$^{1}$, Mark Halpern$^{1}$, Douglas Scott$^{1}$, Gaelen Marsden$^{1}$, Fumihide Iwamuro$^{2}$, Toshinori Maihara$^{3}$, Kentaro Motohara$^{4}$, Tomonori Totani$^{5}$}
\vspace*{6pt}\\
$^{1}$ Department of Physics and Astronomy, University of British Columbia, Vancouver, BC V6T 1Z1, Canada\\
$^{2}$ Department of Physics, Kyoto University, Kitashirakawa, Kyoto 606-8502, Japan\\
$^{3}$ Department of Astronomy, Kyoto University, Kitashirakawa, Kyoto 606-8502, Japan\\
$^{4}$ Institute of Astronomy, University of Tokyo, Mitaka, Tokyo 181-0015, Japan\\
$^{5}$ Princeton University Observatory, Peyton Hall, Princeton, NJ 08544-1001, USA
\vspace*{-0.5cm}
}

\date{Re-submitted June 24 2004}

\maketitle

\begin{abstract}
We have mapped the submillimetre wavelength continuum emission from 
the Subaru Deep Field (SDF) at $450$ and $850\,\mathrm{\mu m}$ with the 
Submillimetre Common-User Bolometer Array (SCUBA) detector on the 
James Clerk Maxwell Telescope (JCMT).  The near-IR image of the SDF is one of the deepest 
near-IR images available and contains four `hyper extremely red objects' 
(HEROs). These data allow us to test the connection 
between `extremely red objects' (EROs) found in IR surveys and the population of 
bright submillimetre sources found with SCUBA.  We present a weak measurement 
of the average flux of the four $\it{K^\prime}$-band selected HEROs of 
$1.15\,(\pm0.46)\,\mathrm{mJy}$, which fails to support the hypothesis 
that HEROs should be bright SCUBA sources.  Our data are consistent with the HEROs being 
objects with SEDs like that of Arp220 out to $z\sim1.7$, however, the extinction in the HEROs must be about 1 magnitude greater in the $\it{J}$-band than is the case for Arp220 and they would need to be 1.7 times
 as luminous as Arp220.  On the other hand, an 
evolutionary model of elliptical galaxies at $z\sim2$--3 in a dusty starburst phase is also in agreement 
with the submillimetre data, as was originally proposed for the HEROs.
\end{abstract}

\begin{keywords}
submillimetre -- cosmology: observations -- galaxies: high-redshift -- galaxies: 
evolution -- galaxies: formation -- galaxies:  starburst
%\vspace*{-0.25cm}
\end{keywords}

\section{Introduction}
\label{sec:intro}
Extremely Red Objects (EROs) emerge at faint near-IR magnitudes 
(${\it{K}} \gtrsim 18$) (\citealt{Elston}; \citealt{McCarthy}; \citealt{Hu})
and have become important in studying the history of galaxy formation and star formation in the
Universe \citep{Yahata,Dickinson}.  
The ERO class contains objects with very red optical to near-IR colours, 
typically ${\it{I-K}} > 4$ or ${\it{R-K}} > 5-6$.  This extreme observed colour can
be produced by old stellar populations at $z \sim 1$ or young 
dust-enshrouded starburst galaxies at high redshifts.  Spectroscopic observations 
of EROs have revealed that about half of 
the population is comprised of very dusty galaxies undergoing 
their initial burst of star formation, while the remaining EROs are evolved stellar populations (\citealt{Dunlop96}; \citealt{Graham}; \citealt{Cohen}; \citealt{Cimatti}).  

Strong submillimetre emission from an ERO is an indication of starburst activity 
(see \citet{Webb}).  In fact,
many of the most luminous submillimetre sources discovered (with typical fluxes of 
$5-20\,\mathrm{mJy}$ at $850\,\mathrm{\mu m}$) have been 
identified as counterparts to EROs (e.g. CUDSS14A \citep{Gear}, 
SMM J00266+1708 \citep{Frayer}, W-MMD11 \citep{Chapman}, SMM J09429+4658, 
and SMM J04431+0210 \citep{Smail}).  Because of the high frequency with which very bright
submillimetre sources are identified with EROs, \citet{Smail} suggest
that SCUBA sources comprise the majority of the reddest EROs.  This would imply that 
a $\it{K}$-band source which is very faint or undetected in optical data (i.e.~$\it{J}$ or $\it{I}$-band) 
would likely be a bright submillimetre source.  Here, we test if the most extremely red objects found to date are 
strong submillimetre emitters.

The Subaru Telescope has been used to produce deep images at {\it{J}}
($\lambda=1.16$--$1.32\,\mathrm{\mu m}$) and $\it{K^{\prime}}$ 
($\lambda=1.96$--$2.30\,\mathrm{\mu m}$) of a `blank' 2 arcmin $\times$ 
2 arcmin field near the north Galactic pole with $5\,\sigma$ detection limits 
of 25.1 and 23.5 in $\it{J}$ and $\it{K^\prime}$, respectively \citep{Maihara}.  
Among the roughly 350 galaxies detected in the $\it{K^{\prime}}$-band in 
the Subaru Deep Field (SDF), there are 4 objects with extreme $\it{J-K^{\prime}}$ 
colours \citep{Maihara}.  These 4 objects have a $1/\mathrm{error}^{2}$ weighted average ${K^\prime}$ 
magnitude and $\it{J-K^{\prime}}$ colour of 21.6 and 3.0, respectively (see Table~\ref{tab:K}).

\begin{table*}
\begin{tabular}{lllll}
\hline
\multicolumn{1}{c}{ID} &
\multicolumn{2}{c}{Position (2000.0)} & 
\multicolumn{1}{c}{$K^{\prime}$} & 
\multicolumn{1}{c}{$\it{J-K^{\prime}}$} \\\hline
SDF1 & \(13^{\mathrm{h}}24^{\mathrm{m}}22{\mathrm{\fs}}38\) & \(+27^{\circ}29'49{\farcs}5\) & 20.91 ($\pm$ 0.05) & 2.97 ($\pm$ 0.14) \\
SDF2 & \(13^{\mathrm{h}}24^{\mathrm{m}}22{\mathrm{\fs}}39\) & \(+27^{\circ}29'01{\farcs}9\) & 22.03 ($\pm$ 0.09) & 3.65 ($\pm$ 0.40) \\
SDF3 & \(13^{\mathrm{h}}24^{\mathrm{m}}21{\mathrm{\fs}}16\) & \(+27^{\circ}29'01{\farcs}9\) & 21.99 ($\pm$ 0.05) & 2.81 ($\pm$ 0.20) \\
SDF4 & \(13^{\mathrm{h}}24^{\mathrm{m}}22{\mathrm{\fs}}84\) & \(+27^{\circ}30'08{\farcs}4\) & 22.31 ($\pm$ 0.14) & 4.12 ($\pm$ 1.04) \\
\end{tabular}
\caption{The positions, $K^{\prime}$ magnitudes, and colours of the HEROs from \citet{Maihara}.  Astrometry is derived from the coordinates of an \textit{HST} guide star in the flanking field and the estimated positional accuracy is $\pm 0{\farcs}15$. }
\label{tab:K}
\end{table*}
 
Among the EROs which are faintest in the $\it{K^\prime}$-band ($K^\prime>22$), an increasing
fraction are extraordinarily red, even compared to other EROs, with
${\it{J-K^{\prime}}}>3{\rm ~or~}4$.  \citet{Totani} call these objects Hyper-EROs
(HEROs) and, in detailed modelling, find that they are too red to be
passively evolving elliptical galaxies.  The HEROs are most likely 
to be either very dusty galaxies which formed at redshifts of $z\approx 4-7$ 
and are still undergoing rapid star formation when seen near 
$z\approx 3$, or they are clean `Lyman-dropout' types of
galaxies seen at extraordinary redshifts of $z\approx 8-10$.  $2\,\mathrm{mJy}$ SCUBA sources have a surface density of $3 \times 10^{3}\mathrm{deg}^{-2}$ \citep{Borys2}, which is similar to the SDF HERO surface density of $\approx 3 \times 10^{3}\mathrm{deg}^{-2}$.  If HEROs comprise the bulk of the SCUBA population, then their number density implies a brightness of $2\,\mathrm{mJy}$.  A measurement of the submillimetre flux of these HEROs would allow us to distinguish between these models.

One of the bright SDF HEROs may be an apparent interacting pair of galaxies (see fig.~\,13 (SDF1) of \citealt{Maihara}, where both components are clearly undetected in the \textit{J}-band), which could be exciting, since mergers have been known to play a role in models of 
ultraluminous infrared galaxies (ULIRGs), which are probably related to SCUBA sources.  Using Poisson statistics and $N\simeq10^{5}\,\mathrm{deg^{-2}}$ down to $\it{K}$$\approx23$ \citep{Maihara}, we estimate that there is only a 5 per cent probability that two galaxies would appear 1.5 arcseconds apart by chance (ignoring clustering).  And with the previous result, we determine that there is an 80 per cent chance of not seeing one galaxy pair (within 1.5 arcsec) among four galaxies.  However, for a HERO surface density of $4.5\times10^{3}\,\mathrm{deg^{-2}}$, the chance that two HEROs in the SDF would lie within 1.5 arcsec of each other at random is 0.25 per cent.  It is therefore quite unlikely to see a pair of HEROs 1.5 arcsec apart if they were not interacting.  One pair of HEROs in the SDF corresponds to $900\,\mathrm{deg^{-2}}$.  How bright do we expect a pair of HEROs to be in the submillimetre?  If they represent a large fraction of bright submillimetre sources, then their surface density suggests a flux density of around $10-20\,\mathrm{mJy}$ (see for e.g. \citealt{Borys2}).

The paper will be organized as follows.  In Section~\ref{obs} we 
discuss the observations, data reduction and fluxes.  The SED galaxy templates and the expected 
$850\,\mathrm{\mu m}$ flux versus redshift relation are discussed in Section~\ref{model}.  
Section~\ref{discussion} presents the results and discussion, and 
Section~\ref{conc} states the conclusions.
Throughout this paper we assume standard cosmological parameter values of 
\(\Omega_{\Lambda}=0.7\), \(\Omega_{\mathrm{M}}=0.3\) and \(H_{\mathrm{0}}=75\,\mathrm{km}\,\mathrm{s^{-1}}\,\mathrm{Mpc^{-1}}\).

\section{Observations, Data Reduction and Fluxes}\label{obs}
The SDF was observed with a resolution of 14.8 arcsec and 7.5 
arcsec, at 850 and $450\,\mathrm{\mu m}$, respectively, with the SCUBA instrument \citep{Holland} on the 
15-m JCMT atop Mauna Kea in Hawaii.  Observations were carried out during May of 2001, May 2002, 
February 2003, and March 2003.  The atmospheric zenith opacity at 
$225\,\mathrm{GHz}$ was monitored with the Caltech Submillimetre Observatory (CSO) tau monitor in 2001 
and 2002 and with the JCMT water vapour monitor in 2003.

One shift of `jiggle mapping' was performed in May 2001 in unstable medium-grade weather, where the 
$850\,\mathrm{\mu m}$ atmospheric zenith opacity ranged from 0.26 to 0.44. Jiggle mapping, centred on an RA and Dec (J2000) of \(13^{\mathrm{h}}24^{\mathrm{m}}21{\mathrm{\fs}}20\) and \(+27^{\circ}29'25{\farcs}0\), was performed using offset pointing centres and the secondary mirror was chopped at a standard frequency of $7.8125\,\mathrm{Hz}$ in RA/Dec coordinates, with chop throws of 30 and 40 arcsec.  Additional jiggle map data were obtained in March 2003 in dry weather, using chops in azimuth in order to reduce the effect of rapid sky variations with alternating position angles of 0 and 90 and chop thows of 30 and 40 arcsec.  Pointing checks were performed hourly on blazars and planets and offsets from the pointing centre were typically 2--3 arcseconds.  

In addition, 9-point photometry data were obtained in May 2002, February 2003, and March 2003, using the central bolometer on the SCUBA array for each of the four HERO sources.  Chop throws were selected to avoid the other sources in the field.  In total, 13.5 and 16.5 hours of useful integration time were obtained for the SDF at $450$ and $850\,\mathrm{\mu m}$ respectively.  

We used SURF (SCUBA User Reduction Facility; \citealt{Jenness}) together with locally developed code, written by \citet{Borys2002}, for data reduction.  The reduction steps were standard, except that the flux from the negative off-beams were then `folded in' to increase the overall sensitivity of the $850\mathrm{\mu m}$ map by a factor of roughly $\sqrt{3/2}$.  We did not do this for the $450\mathrm{\mu m}$ map, since the beam is not sufficiently Gaussian-shaped for this procedure to be simple.

Calibration data were reduced in the same way as the SDF data.  Due to the poor weather and time constraints, few calibrators were observed during the photometry run, and a standard flux conversion factor was adopted for those data.  Neglecting undersampled and noisier map edges, the maps have mean values of $-0.05\,(\pm0.03)\,\mathrm{mJy}$ and $0.3\,(\pm0.2)\,\mathrm{mJy}$ at $850$ and $450\,\mathrm{\mu m}$, respectively.  This is consistent with the expected map average of zero for differential measurements.  The rms values over the maps are about $2.3\,\mathrm{mJy}$ and $11.4\,\mathrm{mJy}$ at 850 and $450\,\mathrm{\mu m}$, respectively.

The flux measurements for the HEROs also include the photometry observations, which are treated as undersampled maps and folded in with the jiggle map data.  Since our targets are unresolved by the JCMT, we simply measure the flux on the beam-convolved SCUBA map at the $\it{K^{\prime}}$-selected position of each HERO.  We then use the noise map to find the noise associated with each pixel, ignoring the calibration uncertainty, which is unimportant at these low signal-to-noise ratios.  The galactic cirrus contribution is $0.9\,\mathrm{MJy\,sr^{-1}}$ (extracted from the \citealt{Schlegel} map) which is much less than the confusion limit of the map.
 
The results are summarized in Table~\ref{tab:flux}.  We fail to significantly detect any of the HEROs and therefore conclude that they are not luminous SCUBA sources.  Since we cannot claim detection of any of our faint sources, Bayesian 95 per cent upper confidence limit flux estimates are calculated for each source by integrating over the non-negative flux regions of a Gaussian probability function.  Combining the results from the four sources allows us to obtain an average object flux for our sample of HEROs.  The error-weighted average flux density for the whole sample at $850\,\mathrm{\mu m}$ is $1.2\,(\pm0.5)\,\mathrm{mJy}$, and has a 95 per cent upper confidence limit of $2.0\,\mathrm{mJy}$.  The average flux density at $450\,\mathrm{\mu m}$ is $0.4\,(\pm3.1)\,\mathrm{mJy}$.

If HEROs comprise a large fraction of the submillimetre source counts of \citet{Borys2}, 
the measured mean brightness of $1.15\,(\pm 0.46)\,\mathrm{mJy}$ would
correspond to sources with a mean density of $6.7\,(\pm 3)$ sources per
$2\,\mathrm{arcmin} \times 2\,\mathrm{arcmin}$ area of sky, 
consistent with the 4 sources found in the Subaru Deep Field.

\begin{table}
\begin{tabular}{lrrc}
\hline
\multicolumn{1}{c}{ID} & \multicolumn{1}{c}{$S_{450}$} & \multicolumn{1}{c}{$S_{850}$} &
\multicolumn{1}{c}{$S_{850}$ 95\% limits}\\
\multicolumn{1}{c}{} & \multicolumn{1}{c}{(mJy)} & \multicolumn{1}{c}{(mJy)} &
\multicolumn{1}{c}{(mJy)}\\\hline
SDF1 & $1.5$ ($\pm$ 5.9) & $2.1$ ($\pm$ 0.9) & $0.4 \leq S_{850} \leq 3.8$\\
SDF2 & $-7.8$ ($\pm$ 6.4) & $1.0$ ($\pm$ 0.9) & $0.0 \leq S_{850} \leq 2.5$\\
SDF3 & $6.1$ ($\pm$ 6.5) & $-0.22$ ($\pm$ 0.9) & $S_{850} \leq 1.6$\\
SDF4 & $1.7$ ($\pm$ 5.8) & $1.8$ ($\pm$ 1.0) & $0.1 \leq S_{850} \leq 3.5$\\
\\
$\bf{Mean}$ & $0.42$ ($\pm$ 3.06) & $1.15$ ($\pm$0.46) & $0.3 \leq S_{850} \leq 2.0$\\
\end{tabular}
\caption{The measured $450$ and $850\,\mathrm{\mu m}$ flux densities and errors for the four SDF HERO sources.  The fourth column lists the 95 per cent Bayesian lower and upper limits to the $850\,\mathrm{\mu m}$ flux.  The bottom row lists the error-weighted mean of the $850\,\mathrm{\mu m}$ fluxes and the 95 per cent Bayesian lower and upper confidence limit to this mean flux.  The 95 per cent Bayesian upper confidence limit to the $450\,\mathrm{\mu m}$ flux is $\leq 6.3\,\mathrm{mJy}$.}
\label{tab:flux}
\end{table}

\section{The Model}\label{model}
In order to interpret our results, we need a realistic model to describe expectations for the submillimetre properties of the HEROs.  We use a phenomenological model, based on scaling the properties of known prototypes.   We utilize the $850\,\mathrm{\mu m}$ fluxes, $\it{K^{\prime}}$ magnitudes and near-IR colours, to fit a three parameter model to a redshifted SED model.  The model parameters include the overall luminosity, the amount of dust reddening in the IR, and the redshift.

\subsection{ULIRG model}
We created a template SED using public photometric data available for Arp 220 from the NASA/IPAC Extragalactic Database (NED).  Arp 220 is the most luminous object in the nearby Universe (\(z=0.018)\) and is a well-studied example of a ULIRG.  Higher redshift sources have more sparsely sampled SEDs, and so in using Arp220, we start at least on firm ground.  

\begin{figure}
\epsfig{file=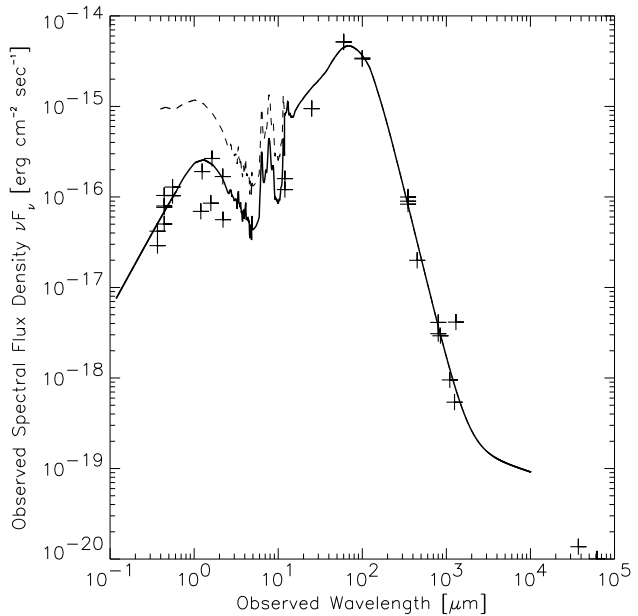,width=0.488\textwidth}
\caption{The model spectral energy distribution of a nearby ULIRG: the \citet{Lagache} starburst template (dashed line represents number 30 of 41 possible templates) fit at $z=0.018$ to Arp220 photometric data (solid line is an extinction corrected version), which are marked with the plus signs.  The slope of the drop at $\lesssim 1\,\mathrm{\mu m}$ has been extrapolated to shorter wavelengths.}
\label{fig:sed}
\end{figure}

We then use the \citet{Lagache} UV--to--radio SED models of a typical ULIRG to interpolate the Arp220 photometric points (see Fig.~\ref{fig:sed}).  These templates evolve in luminosity (i.e. become hotter) and cover a wide range of luminosity levels from $\mathit{L}=10^{9}-10^{13}\mathrm{L_{\sun}}$.  We employ their model number 30.  We increase the template's IR dust content (see details below) since the \citet{Lagache} starburst template is not red enough to fit the Arp220 photometric points.  In addition, incorporating a dust model provides a mechanism for varying the amount of dust extinction to fit the HERO colours.

We redden the original template by multiplying it by an extinction function which varies with wavelength:
\begin{equation}
f = \left(1 + \left(\frac{1.23}{\lambda}\right)^{2.6}\right)^{-0.2x},
\label{eqn:dust}
\end{equation}
where the free parameter, \(x\), is adjusted to a value of $x=2.5$ in order to match the NED data points, and \(\lambda\) is wavelength in $\mathrm{\mu m}$.  The above extinction law was constructed using a power-law fit to the relative extinction in the Landolt $\it{V}$, $\it{R}$, $\it{I}$ and UKIRT $\it{J}$, $\it{H}$, $\it{K}$ and $\it{L^{\prime}}$ bandpasses versus effective wavelength from \citet{Schlegel}.  They employ a `diffuse ISM' mean value of $R_{V}=3.1$ for the extinction laws of \citet{Cardelli} and \citet{O'Donnell}.  We note that the difference between the extinction curves of the Milky Way, the Magellanic Clouds and starburst galaxies is almost negligible at wavelengths longer than $\approx0.26\,\mathrm{\mu m}$ (see \citealt{Calzetti, Cardelli}).  Hence variations are unimportant for $z<3$ galaxies (where the $\it{J}$-band corresponds to rest wavelengths of $\lesssim 0.31\,\mathrm{\mu m}$).

\subsection{Spiral Galaxy Model}
\begin{figure}
\epsfig{file=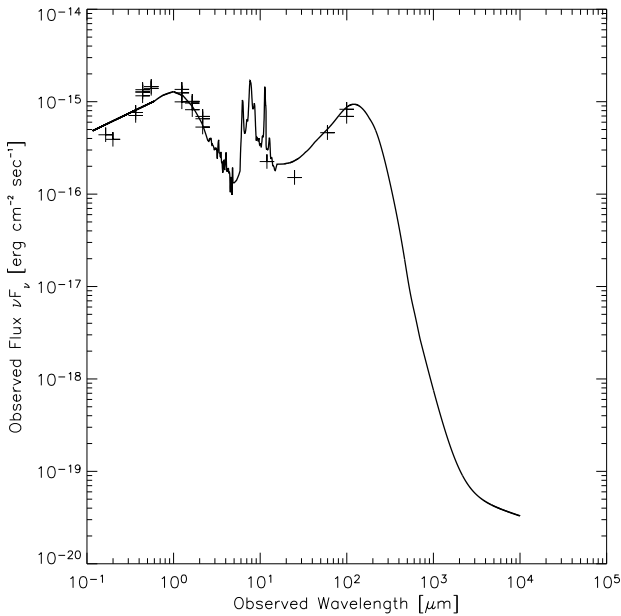,width=0.488\textwidth}
\caption{The model spectral energy distribution of a nearby normal spiral galaxy: the \citet{Lagache} normal galaxy template fit at $z=0.002699$ to NGC3938 photometric data, which are marked with the plus signs.  Again, the model SED has been extrapolated to shorter wavelengths.}
\label{fig:sed_norm}
\end{figure}

For comparison, we also created a normal spiral galaxy SED using NED public photometric data for NGC3938.  NGC3938 is a well-studied, multiple-armed, early luminosity class Sc(s) face-on galaxy $10.8\,\mathrm{Mpc}$ away.  We utilize the \citet{Lagache} template of a normal cold spiral galaxy scaled evenly across all wavelengths with no addition of dust required (see Fig.~\ref{fig:sed_norm}).

\subsection{Using the Templates}
We have constructed three-parameter models based on these SEDs, constrained only by the ${K^\prime}$ magnitude and the $\it{J-K^{\prime}}$ colour.  We choose a redshift, redden the template to match the observed $\it{J-K^{\prime}}$ colour by varying $x$ in Eqn.~\ref{eqn:dust}, then boost or diminish the template achromatically to match the $\it{K^{\prime}}$ magnitude.

We note that performing numerical integration over the Arp220 SED of Fig.~\ref{fig:sed} reveals that the total power absorbed in reddening the starburst galaxy template is only about $3-5$ per cent of the total galactic emission.  The small increase in the far-IR luminosity can be neglected as the reddening is altered.  In contrast, Fig.~\ref{fig:sed_norm} reveals the energetics of the normal cold spiral galaxy, and it is clear that there is just as much power in the near-IR peak as in the far-IR peak.  Therefore, the power that would be `lost' due to reddening the spiral galaxy must be added back into the far-IR region of the spectrum in order to conserve energy.  We adjust the FIR luminosity accordingly in order to account for the power absorbed in reddening the galaxy.

We scale the template brightness uniformly across each spectrum to fit the inferred ${K^\prime}$ flux, as determined from the observed ${K^\prime}$ magnitude using:
\begin{equation}
	F_{\nu}=10^{-0.4m}F_{0},
\end{equation}
where $\it{m}$ is the magnitude and $F_{0}$ is the zero-point flux which is \(718.9\,\mathrm{Jy}\) for the $\it{K'}$-band, and was extracted from the interpolation of the broad-band flux of Vega.

\begin{figure}
\epsfig{file=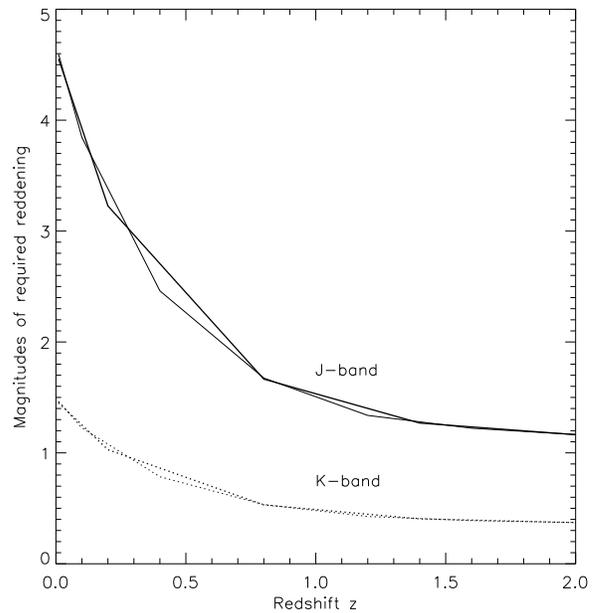,width=0.488\textwidth}
\caption{The amount of dust extinction that we must add to our modified Arp220 SED (bold lines) and to our normal spiral galaxy SED for increasing redshift at the $\it{J}$ (solid lines) and $\it{K^\prime}$ (dotted lines) bands to achieve the mean observed $\it{J-K^{\prime}}$ colour.}
\label{fig:dust}
\end{figure}

\begin{figure}
\epsfig{file=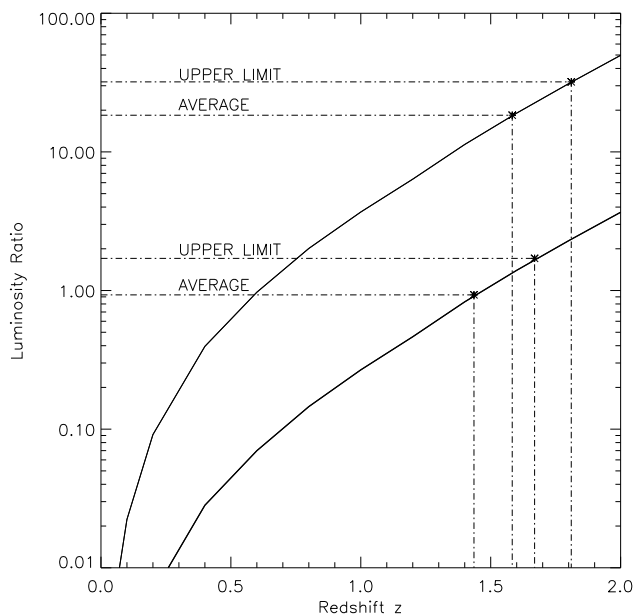,width=0.488\textwidth}
\caption{A plot of the brightness in the $\it{K^{\prime}}$-band (giving the correct $\it{K^{\prime}}$ magnitude) compared to the brightness of Arp220 (lower bold line) and a normal spiral galaxy (upper line) after adding in the appropriate amount of extra dust to obtain the observed $\it{J-K^{\prime}}$ colour, plotted versus redshift for our mean HERO.}
\label{fig:bright}
\end{figure}

\begin{figure}
\epsfig{file=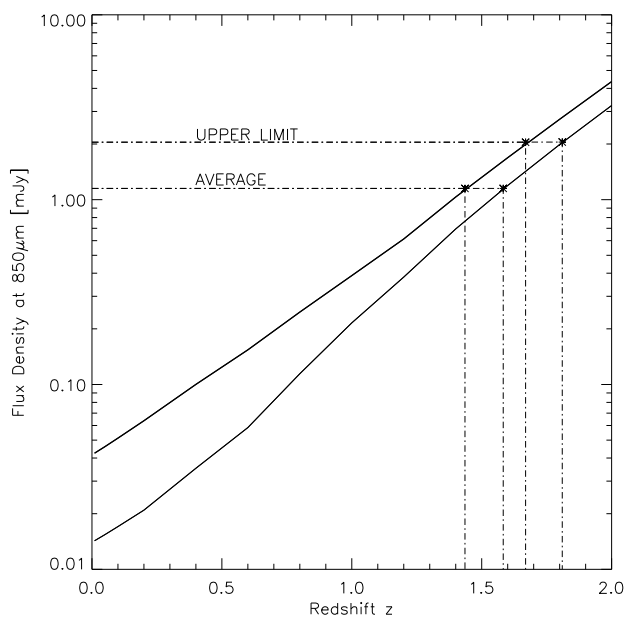,width=0.488\textwidth}
\caption{Expected $850\,\mathrm{\mu m}$ flux density versus redshift for our mean HERO assuming a modified Arp220 SED (upper bold line) and a normal spiral galaxy (lower line).  Asterisks represent where we read off a redshift, based on the mean and upper limit fluxes.}
\label{fig:all}
\end{figure}

After fitting the template to the ${K^\prime}$ magnitude and $\it{J-K^{\prime}}$ colour, we then read off the observed $850\,\mathrm{\mu m}$ flux.  We repeat this process for a range of redshifts (see Fig.~\ref{fig:all}).  Added dust in the $\it{J}$ and $\it{K^{\prime}}$ bands versus redshift is plotted in Fig.~\ref{fig:dust}.  We plot the scale factor versus redshift in Fig.~\ref{fig:bright}.

\section{Results and Discussion}\label{discussion}
We estimate a redshift for each HERO using the recipe that follows.  At each redshift, we redden the template to produce the observed $\it{J}-\it{K^{\prime}}$ colour of each HERO.  Next, we scale the whole SED in order to match the observed $\it{K^{\prime}}$ magnitude.  We then test whether the predicted $850\,\mathrm{\mu m}$ flux is consistent with the observed value.  Figs.~\ref{fig:dust}, \ref{fig:bright} and \ref{fig:all} can be read to find the properties of a ULIRG or a spiral galaxy corresponding to our measurements.

\subsection{Could the HEROs be Starburst Galaxies?}
Using Fig.~\ref{fig:all} we see that an object with an SED like that of Arp220 and properties
corresponding to the mean of our measurements in Table~\ref{tab:flux} must be
located at $z=1.44\,(\pm0.10)$.  From Fig.~\ref{fig:bright}, one sees that such a reddened
object at $z=1.44$ must be $0.9\,(\pm0.2)$ times as luminous as Arp220 to
match the observed $\it{K^{\prime}}$-band flux, and from Fig.~\ref{fig:dust}, the amount of
reddening in $\it{K^{\prime}}_{\mathrm{observed}}$ is 0.4 magnitudes compared to Arp220.  
Our observed 95 per cent upper confidence limit for the four HEROs of $S_{850}\leq
2.0\,\mathrm{mJy}$ leads in the same way to $z \leq 1.7$ (see Table~\ref{tab:z}), $L_{\nu\mathrm{(HERO)}}/L_{\nu\mathrm{(ULIRG)}}\leq 1.7$, and $\Delta K \geq 0.4$ magnitudes of extinction.  This is consistent with the median
redshift estimates for SCUBA sources from \citet{Dunlop} and \citet{Chapman3}, but it is at the low redshift end.  If the HEROs have SEDs like Arp220 and are located at the mean redshift reported by \citet{Chapman3} of $z=2.4$ and possess the observed $\it{J}$ and $\it{K^{\prime}}$-band fluxes,
one would expect them to be $12\,\mathrm{mJy}$ sources, which they are not.

\begin{table}
\begin{tabular}{lrrrrr}
\hline
\multicolumn{1}{c}{ID} & \multicolumn{2}{c}{Upper Limits}\\
\multicolumn{1}{c}{} & \multicolumn{1}{c}{$z_\mathrm{ULIRG}$} & \multicolumn{1}{c}{$z_\mathrm{spiral}$}\\\hline
SDF1 & $< 1.7$ & $< 1.8$ \\
SDF2 & $< 1.6$ & $< 1.7$ \\
SDF3 & $< 1.8$ & $< 2.0$ \\
SDF4 & $< 1.6$ & $< 1.8$ \\
\\
$\bf{Mean}$ & $< 1.7$ & $< 1.8$ \\
\end{tabular}
\caption{Model-estimated redshifts (for the ULIRG and the normal spiral galaxy) based on the $850\,\mathrm{\mu m}$ upper limit flux of each source.  The bottom row lists the model-estimated redshift based on the upper limit flux of our mean HERO.}
\label{tab:z}
\end{table}

In principle the 95 per cent upper confidence limit of $S_{450} \leq 6.3\,\mathrm{mJy}$ provides a redshift upper limit of $z\leq 1.44$, which is tighter than the limit obtained from the $850\,\mathrm{\mu m}$ data.  However, the value of this limit relies on the precise understanding of the $450\,\mathrm{\mu m}$ variance and we therefore do not rely on it.

We have tested our Arp220 model by comparing it to those
very few objects for which IR colour, $\it{K}$-band magnitude, $850\,\mathrm{\mu m}$ flux and
redshift are all known.  We use an ERO and a VRO (very red object) from the Hubble Deep Field (HDF) \citep{Borys2} with measured spectroscopic redshifts.  We also use HDF850.1 \citep{Hughes}, a SCUBA-bright IR-faint ERO with a secure photometric redshift \citep{Dunlop2002}.  We also include HR10 (or ERO J164502+4626.4, \citealt{Dey}), the first ERO discovered to be associated with a ULIRG, and FN1 40 from the Far InfraRed Background (FIRBACK) survey (\citealt{Dolethesis}, \citealt{Sajina}).  See Table~\ref{tab:test} for the IR colours, submillimetre fluxes, spectroscopic redshifts and model-predicted redshifts.  The mean difference between the spectroscopic redshifts and the model-estimated redshifts ($z_{\mathrm{model}}-z_{\mathrm{spec}}$) is $+0.06$ with a spread of $\pm\sigma=0.6$.  We are therefore confident that this simple model is suitable for estimating redshifts of the reddest population of objects when the IR colour and the $850\,\mathrm{\mu m}$ flux are known.  The model possesses an advantage over more complicated multi-parameter models in that it fits known sources well, despite its single-parameter simplicity.

\begin{table*}
\begin{tabular}{llllll}
\hline
\multicolumn{1}{c}{ID} & \multicolumn{2}{c}{Optical} & \multicolumn{1}{c}{$S_{850}$} & \multicolumn{1}{l}{$z_{\mathrm{spec}}$} & \multicolumn{1}{l}{$z_{\mathrm{model}}$}\\
\multicolumn{1}{c}{} & \multicolumn{1}{c}{\textit{I}} & \multicolumn{1}{c}{\textit{I-K}} & \multicolumn{1}{c}{(mJy)} & \multicolumn{1}{c}{} & \multicolumn{1}{c}{}\\\hline
SMMJ123628+621048 & $22.5$ & $4.0$ & $4.4\pm1.2$ & 1.013 & $1.2\pm0.3$\\
SMMJ123635+621239 & $22.3$ & $3.5$ & $3.0\pm0.8$ & 1.219 & $1.3\pm0.3$\\
SMMJ123652+621227 & $28.7$ & $5.2$ & $7.0\pm0.5$ & $4.1\pm0.5$ & $3.4\pm0.1$\\
SMMJ164502+4626.4 & $24.9$ & $6.5$ & $4.9\pm0.7$ & 1.44 & $0.82\pm0.1$\\
FIRBACK FN1 40 & $23.96$ & $4.23$ & $6.3\pm1.4$ & $0.449$ & $1.8\pm0.2$\\
\end{tabular}
\caption[Submillimetre sources with red IR colours and known redshifts]{Submillimetre sources with red IR colours and known redshifts.  Model-predicted redshifts and errors (estimated using the errors in the $850\,\mathrm{\mu m}$ fluxes) are given in the last column.}
\label{tab:test}
\end{table*}

Since the HEROs are extreme examples of EROs, it is useful to consider how the results would change by fitting an ERO SED to the data.  HR10 \citep{Hu} is an archetypal ULIRG-ERO association and is merely a distant clone of Arp220 ($z=1.44$) and 3.8 times more luminous \citep{Elbaz}.  We construct a template of HR10 and find nearly identical constraints on redshift and luminosity.

 We treat the model of \citet{Totani} (see \citealt{TotaniTak} for details of the model) 
of proto-elliptical galaxies which
formed at $z=3$ and are observed at $z=2.3$ as a template for emission.  We apply Eqn.~\ref{eqn:dust} to redden the template and we find that achieving a colour of $\it{J}-\it{K^{\prime}}=$ 4.12 requires 1.80 magnitudes of extinction in \textit{J} and 0.57 magnitudes of extinction in $\it{K^{\prime}}$.  The resulting $850\,\mathrm{\mu m}$ flux is $0.5\,\mathrm{mJy}$.  This is close to the $1.0\,\mathrm{mJy}$ prediction from \citet{Totani} and consistent with our measured average flux
of $1.15\,(\pm 0.46)\,\mathrm{mJy}$.  However, there are $\sim 2\times 10^{4}$ $0.5\,\mathrm{mJy}$
SCUBA sources per $\mathrm{deg}^{2}$, according to the number counts of \citet{Borys2}.  Recall that the surface density of HEROs in the SDF is $\sim 3\times 10^{3}\,\mathrm{deg}^{-2}$, which corresponds to the surface density of $2\,\mathrm{mJy}$ SCUBA sources.  If HEROs are typical elliptical
galaxies then they only comprise a small fraction of SCUBA-detected sources.

\subsection{Could the HEROs be Normal Spiral Galaxies?}
We now turn to the normal spiral galaxy SED.  When we force our template SED to match the 95 per cent upper limit to the $850\,\mathrm{\mu m}$ flux of $2.0\,\mathrm{mJy}$, constrained by a faint $\it{K^{\prime}}$ magnitude and very red $\it{J-K^{\prime}}$ colour, we estimate a redshift of $z<1.8$ (see Fig.~\ref{fig:all}).

At $z=1.8$, the SED requires a factor of $\simeq32$ in luminosity boosting in order to reproduce the correct $\it{K^{\prime}}$ magnitude for the HERO sample (see Fig.~\ref{fig:bright}).  At $z=1.8$, we require $\approx0.4$ magnitudes of extinction in $\it{K^\prime}$ and $\approx1.2$ magnitudes of extinction in {\it{J}} (see Fig.~\ref{fig:dust}).

Because of the large luminosity boosting required, NGC3938 is not a reasonable analogue for HEROs if they are near to the $850\,\mathrm{\mu m}$ flux upper limit.  However, if the HEROs are normal galaxies at a redshift near to 0.6 and reddened by 2 magnitudes in $\it{J}$-band, they would produce
the observed $\it{J}$ and $\it{K^\prime}$-band fluxes with no luminosity boosting required. The $850\,\mathrm{\mu m}$ flux would be $0.06\,\mathrm{mJy}$.  This is below the 95 per cent lower confidence limit of $S_{850} \geq 0.3\,\mathrm{mJy}$, although the value of this constraint depends on the reliability of our estimate of the variance in the data.

\subsection{Other Possibilities}
Can the HEROs be evolved elliptical galaxies at higher redshifts?  We know that the reddest and faintest EROs at known redshifts are not elliptical galaxies.  Therefore, the HEROs are most likely not well-described by high-redshift passively evolving elliptical galaxies that are red due to old stellar populations.

If the HEROs are \textit{J}-band dropout galaxies, they must lie at redshifts of $z\approx8-10$.  \citet{Totani} state that the UV luminosity inferred from the $\it{K^{\prime}}$ magnitudes of the HEROs implies star formation rates of $\gtrsim 140\,M_{\odot}\,\mathrm{yr^{-1}}$, which sets a total mass of $M\gtrsim10^{11}\,M_{\odot}$ for each system.  They find that the calculated surface density of this population of massive objects at such early times (in the standard theory of CDM hierarchical structure formation) is much too small compared with the observed number density of HEROs in the SDF.  Thus, there is no reason to suspect that these objects are $\it{J}$-band dropout galaxies at extraordinary redshifts.

%The general trend of the model, that the higher redshift objects are
%dustier and more luminous, is consistent with \citet{Blain}.  One
%expects 700 submillimetre sources per $\mathrm{deg}^{2}$ brighter 
%than $5\,\mathrm{mJy}$ from \citet{Borys2}.  Under the same assumption 
%that HEROs comprise a subset of the
%submillimetre source counts, an IR survey of 15 square arcminutes to
%the depth of the Subaru Deep Field survey would turn up a HERO
%corresponding to an $850\,\mathrm{\mu m}$ source brighter than $5\,\mathrm{mJy}$ 
%with 95 per cent confidence.  Our model indicates sources brighter 
%than $5\,\mathrm{mJy}$ might well be at
%redshifts above $z=2.0$, consistent with the distribution of bright
%submillimetre sources for which redshifts are known.

\section{Conclusions}\label{conc}
We observed four objects with extreme $\it{J-K^{\prime}}$ colours (HEROs) with SCUBA on the JCMT.  We obtained no statistically significant detections of the individual objects, but we obtained a marginally significant detection of the objects as a group, finding a mean $850\,\mathrm{\mu m}$ flux of $1.15\,(\pm0.46)\,\mathrm{mJy}$ or a combined 95 per cent upper confidence limit of $2.0\,\mathrm{mJy}$.  This is no more than a hint that they are dusty galaxies emitting a small amount of submillimetre flux.  

HEROs could have SEDs like that of Arp220, but the ones we examined must reside
at the small redshift end of submillimetre redshift distribution, with
an average redshift of $z\approx 1.4$. These objects would have rest frame luminosities 
comparable to that of Arp220 and would be only slightly reddened.

HEROs could alternately be the progenitors of elliptical galaxies at the
end of a dusty starburst phase, as \citet{Totani} suggest.  In this case, they would be at the dim end of the submillimetre flux range that our measurements allow, and they do not comprise the majority
of $850\,\mathrm{\mu m}$ sources at that flux level.

HEROs are not likely to be ordinary elliptical or spiral galaxies at any redshift.

In order to confirm that HEROs are bona fide SCUBA sources (which requires detecting individual objects with SCUBA), a much larger survey would need to be conducted.  Assuming that the HEROs comprise a subset of the
submillimetre source counts, a near-IR survey of $15\,\mathrm{arcmin^{2}}$ to
the depth of the Subaru Deep Field survey should reveal a HERO
corresponding to an $850\,\mathrm{\mu m}$ source brighter than $5\,\mathrm{mJy}$ with 95 per cent confidence.  Our model indicates that sources brighter than $5\,\mathrm{mJy}$ would be at
redshifts above $z=2.0$, which is consistent with the distribution of bright
submillimetre sources for which redshifts are known.

SCUBA galaxies are very often identified with EROs and we have tested if the inverse statement also holds true.  Our results clearly demonstrate that measuring the submillimetre flux of HEROs will not necessarily select very bright SCUBA sources.  SCUBA galaxies may just comprise a subset of a more diverse population of EROs.

\section{Acknowledgments}\label{ack}
This work was supported by the Natural Sciences and Engineering Research Council of Canada.  We would like to thank the staff of the JCMT for their assistance with the SCUBA observations.  The James Clerk Maxwell Telescope is operated on behalf of the Particle Physics and Astronomy Research Council of the United Kingdom, the Netherlands Organisation for Scientific Research, and the National Research Council of Canada.  This research has made use of the NASA/IPAC Extragalactic Database (NED) which is operated by the Jet Propulsion Laboratory, California Institute of Technology, under contract with the National Aeronautics and Space Administration.  We acknowledge the use of NASA's $\it{SkyView}$ facility (http://skyview.gsfc.nasa.gov) located at NASA Goddard Space Flight Center.  We are grateful to Guilaine Lagache for providing us with her starburst templates and an unpublished version of a paper describing her galaxy templates and to Anna Sajina for help using the templates.  We also wish to thank an anonymous referee for constructive comments on an earlier version of this paper.

\end{document}